# Controlling Switching Evolution in Lead-Free Perovskite-Inspired Chalcogenide Memristors for Neuromorphic Computing

Emmanuel Joseph Shaji[1,2,$], Zhiyuan Li[3,$], Srikanth Doddapaneni[1,2], Bhavya Rakheja[4], Vikrant Chaudhary[3,8], Avantika Suthar[5], Lingyun Zhu[1,2], Jingxin Ma[1,2], Hongbin Zhang[3], Monojit Bag[5], Gerardo Hernandez-Sosa[1,2,6,7], Ramesh Kumar[1,2,*]

[1]Light Technology Institute (LTI), Karlsruhe Institute of Technology, Engesserstrasse 13, 76131 Karlsruhe, Germany.

[2]InnovationLab, Speyerer Strasse 4, 69115 Heidelberg, Germany.

[3]Institute of Materials Science, Technical University of Darmstadt, Darmstadt 64287, Germany

[4] Department of Materials Science and Engineering; Solar Cell Technology, Uppsala University, Uppsala, Box 35 SE-751 03, Sweden.

[5]Advanced Research in Electrochemical Impedance Spectroscopy Laboratory, Indian Institute of Technology Roorkee, Roorkee 247667, India.

[6]Institute for Automation and Applied Informatics, Karlsruhe Institute of Technology, Eggenstein-Leopoldshafen, Germany

[7]Institute of Microstructure Technology, Karlsruhe Institute of Technology (KIT), Eggenstein-Leopoldshafen, Germany

[8]Physics Department and CSMB, Humboldt-Universitat zu Berlin, 12489 Berlin, Germany

[$]Equal contribution

*Ramesh Kumar (ramesh.kumar@kit.edu)

**Abstract**

Memristors have emerged as key building blocks of neuromorphic computing architectures due to their ability to integrate data storage and processing. While metal halide perovskites have recently shown significant promise owing to their mixed electronic-ionic conduction and low-cost solution processability, their reliance on toxic lead and limited stability presents critical challenges. Here, we report environmentally friendly, low-toxicity $AgBiS_2$-based solution-processable memristors exhibiting an ultra-low SET voltage of ~0.08 V and a high ON/OFF ratio of $>10^4$. First-principles calculations identify Ag interstitials as the energetically most favourable native defect and reveal low migration barriers within the Ag sublattice, for both interstitials and vacancies, facilitating ionic transport in the $AgBiS_2$ lattice. Through interface and thickness engineering, the resistive switching behaviour can be systematically tuned from abrupt digital to gradual analog modes. Notably, thicker switching layers promote the evolution of stable conductive pathways through intermediate metastable states, revealing a controllable filament evolution process. Electrochemical impedance spectroscopy reveals pronounced negative capacitance (inductive) behaviour at low bias voltages, arising from coupled electronic-ionic dynamics. Consistent with this behaviour, pulse measurements demonstrate gradual conductance modulation under pulse trains, emulating synaptic responses relevant for neuromorphic computing. Finally, post-operando structural analysis reveals substantial morphological evolution of the switching layer driven by repeated filament formation and rupture. Linking structural dynamics to switching variability provides important design principles for achieving reliable and durable sustainable memristors.

## 1. Introduction

With the rapid emergence of AI across all sectors, there is an increasing need for energy-efficient and faster computer architectures.[1] In conventional computing systems, data storage and processing are separated, which creates a bottleneck known as the "memory wall".[2,3] To overcome this limitation, neuromorphic computing hardware integrates data storage and processing within the same units, mimicking the structure and functionality of the human brain.[4,5] In this context, neuromorphic chips based on traditional complementary metal-oxide-semiconductor (CMOS) technology require a large number of transistors and capacitors to emulate biological processes, resulting in increased architectural complexity, high power consumption, and poor cost efficiency.[6] To address these limitations, memristors have emerged as a key enabling technology due to their ability to combine both volatile and non-volatile retention characteristics, along with support for digital and analog computation.[7,8] These features provide significant advantages, including high integration density and low power consumption, making memristors highly suitable for energy-efficient computing architectures. Memristors are two-terminal devices whose resistance state changes in response to an applied voltage or current and retains memory of its previous state. Among the various material systems, including oxides[9], two-dimensional (2D) materials[10,11], and phase-change materials[12], metal halide perovskites (MHPs) stand out as particularly promising candidates due to their mixed electronic-ionic conduction, superior optoelectronic properties, and low-cost solution processability.[13–15] Since 2009, MHPs have emerged as remarkable materials for optoelectronic applications.[16,17] However, their intrinsic ionic conduction, arising from the soft lattice nature, poses challenges such as instability and anomalous device kinetics.[18–20] These effects manifest as phenomena such as hysteresis in the current-voltage (*I-V*) characteristics and large apparent negative capacitance of solar cells.[21–23] These phenomena have motivated the exploration of MHPs for neuromorphic computing, where their unique properties can be harnessed to emulate synaptic behaviour.[24–26] Furthermore, their optoelectronic characteristics offer additional potential for in-sensor computing.[27] In this direction, recent years have witnessed a significant surge in research efforts.

However, high-performance MHPs are mostly based on lead (Pb), which raises concerns due to its toxicity, and these materials also suffer from limited chemical stability.[28,29] Therefore, alternative mixed electronic-ionic semiconductors have been explored, particularly perovskite-inspired materials (PIMs).[20,30–33] These materials are typically based on elements such as bismuth (Bi)[34] and antimony (Sb)[35], as well as chalcogenide compounds like $AgBiS_2$ and $Ag_2S$, offering a more environmentally friendly and potentially stable platform for device applications.[36,37] Chalcogenide compounds remain largely unexplored for memristor applications.[38] Among them, the ternary compound $AgBiS_2$ exhibits several promising properties, including a tunable band gap, low-cost solution processability, high defect density, and good stability comparable to MHPs and other PIMs.[39,40] Although these materials have been investigated for photovoltaic applications, their performance has remained limited due to high recombination losses arising from defects.[41] However, these defect states, which are detrimental to photovoltaics, could be advantageous for memristor applications.[19,42,43] To the best of current knowledge, only a very limited number of $AgBiS_2$- based memristors have been reported in the literature.[44,45] Several important questions remain unresolved, including the material-level origins of defect formation, possible ion-migration pathways, and their implications for device behaviour and the underlying switching mechanisms. Furthermore, the roles of device interfacial engineering and switching-layer thickness in governing electrical switching and $I$-$V$ characteristics remain largely unexplored. Moreover, the effects of mixed electronic-ionic conduction in these materials on device physics remain largely unexplored. In MHPs and PIMs, such coupled transport has been reported to give rise to anomalous electrical characteristics, including negative capacitance and inductive loops, which can play an important role in the emergence of synaptic properties.[20,26,35] However, whether similar mixed electronic-ionic effects occur in solution-processable perovskite-inspired chalcogenides based memristors, and how they influence their electrical and neuromorphic behaviour, remains unclear.

Here, we report environmentally friendly, low-toxicity $AgBiS_2$ (Ag-Bi)-based solution-processable memristors. Through atomistic simulations, we identify the dominant defect states, their formation energies, and migration pathways within the Ag-Bi lattice, providing insights into the microscopic origins of resistive switching. Furthermore, through device engineering, we systematically vary the switching-layer thickness and introduce interfacial layers to tune the switching behaviour from digital to analog modes. Electrochemical impedance spectroscopy under different bias conditions is employed to elucidate defect dynamics and electronic-ionic coupling mechanisms. Finally, long-term stability and post-cycling structural analyses reveal the role of filament evolution and morphological changes in governing device reliability. These results provide fundamental insights into defect-mediated switching mechanisms and establish $AgBiS_2$ as a promising lead-free platform for energy-efficient memory and neuromorphic computing applications

## 2. Results and Discussion

### 2.1. Crystal Structure and Defect Chemistry of $AgBiS_2$

X-ray diffraction (XRD) measurements of the solution-processed $AgBiS_2$ thin film (**Figure 1b**) reveal diffraction peaks indexed to the (111), (200), (220), and (311) planes, consistent with cubic silver bismuth sulphide (ICDD: 01-089-2045), as shown in **Figure 1a**. To understand the defect chemistry of $AgBiS_2$, first-principles calculations were performed on cation-disordered $AgBiS_2$ using a 64-atom special quasirandom structure (SQS) representative of the disordered rock-salt lattice.[46,47] Identifying the energetically favourable native defects and their migration pathways is important because these defects govern ionic transport and can provide insight into the microscopic origins of the resistive switching mechanisms discussed later. Silver vacancies ($V_{Ag}$) were computed for all sixteen Ag sites of the supercell, while silver interstitials ($Ag_i$) were identified using a Voronoi-based scheme[48,49] and, following geometric screening, the three most open interstitial cavities were retained. Under the chemical potential conditions relevant to device operation, the formation energies of both $Ag_i$ and $V_{Ag}$ are negative, with $Ag_i$ lying ~1.1-1.6 eV below $V_{Ag}$ (**Figure 1c**). These results suggest that Ag Frenkel disorder is energetically favourable within the film itself, providing an intrinsic reservoir of mobile $Ag_i$ species without requiring a continuous supply from the Ag electrode. Regarding charge states, the $\varepsilon(+1/0)$ transition levels of all three $Ag_i$ sites lie above the conduction band minimum (0.77-0.86 eV, compared with a bandgap of 0.69 eV), indicating that $Ag_i$ remains in the +1 charge state throughout the bandgap, whereas the $V_{Ag}$ transition levels are located within the gap (0.12-0.45 eV). Since these defects are generated as a Frenkel pair when an Ag atom leaves its lattice site, the pair must remain charge-compensated; the $Ag_i^+ + V_{Ag}^-$ configuration is calculated to lie 0.32–0.63 eV below the neutral pair for all site combinations considered, consistent with the experimental assignment of $Ag^+$ as the dominant mobile species. For migration, 33 Ag-related pathways involving both $V_{Ag}$ and $Ag_i$ yielded activation barriers ranging from 0.08 to 0.53 eV (median ~ 0.24 eV), as calculated using the nudged elastic band method.[50] In comparison, three $V_{Bi}$ migration pathways exhibit significantly higher barriers of 0.75-0.96 eV (**Figure 1d**). These results suggest that ionic transport is dominated by the Ag sublattice, while displacement of the Bi-S framework is energetically much less favourable. It is also worth noting that none of these quantities is a single number: across sites with different local environments, the formation energies, transition levels and migration barriers are spread over roughly 0.4 - 0.6 eV, 0.2 - 0.3 eV and 0.45 eV, respectively due to the intrinsic cation disorder of $AgBiS_2$.

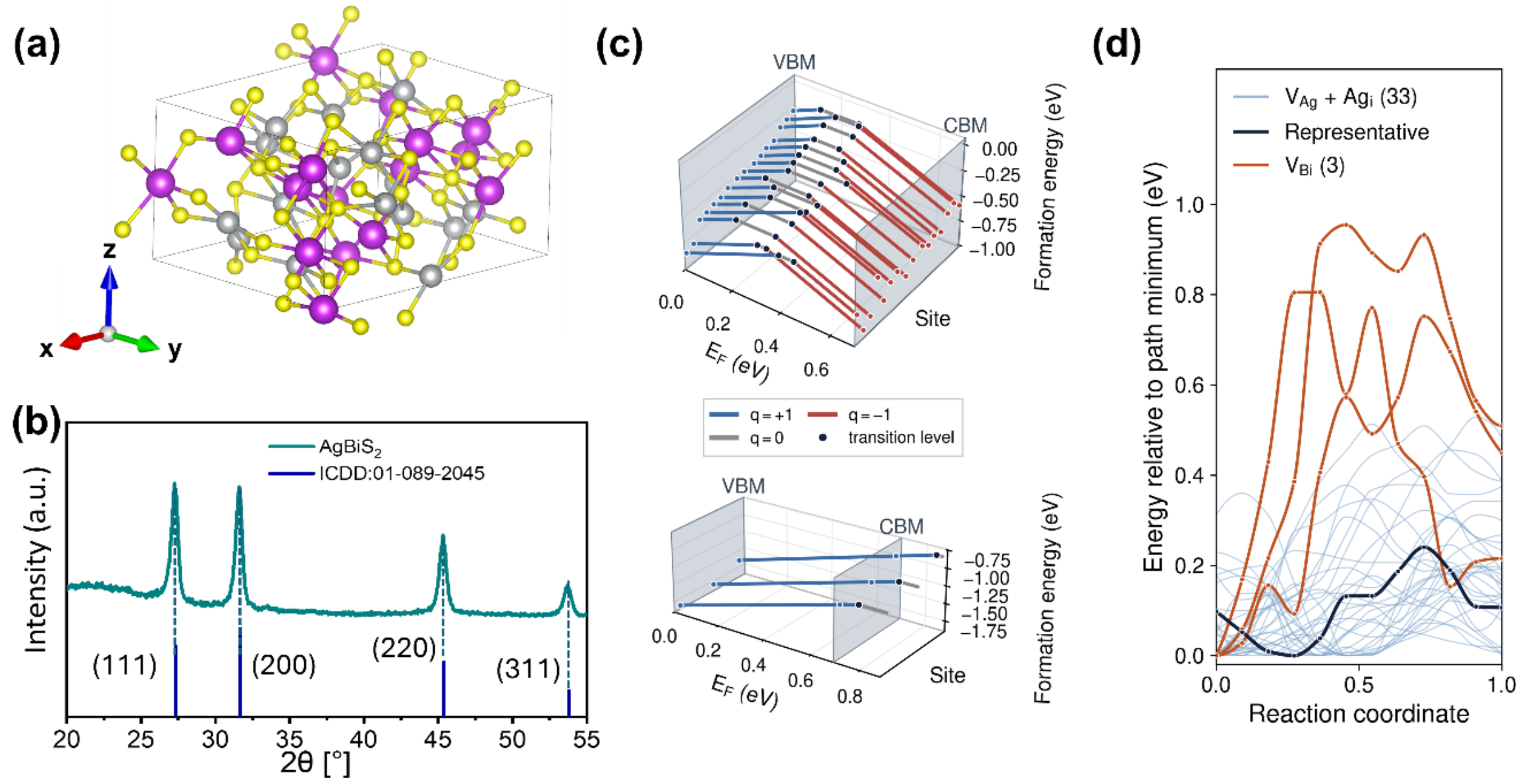

**Figure 1.** (a) Crystal structure of cation-disordered $AgBiS_2$, in which Ag and Bi share the cation sublattice of the rock-salt lattice. (b) X-ray diffraction pattern of the solution-processed $AgBiS_2$ film, with the reflections indexed to the cubic phase (ICDD: 01-089-2045). (c) Calculated formation energies of silver vacancies ($V_{Ag}$, top) and silver interstitials ($Ag_i$, bottom) in cation-disordered $AgBiS_2$, under the chemical potential imposed by the device. Each polyline is the ground-state envelope of a single site, coloured by charge state; its kinks are the charge transition levels (dark points), and the shaded planes mark the VBM and CBM. Sites are ordered by their formation energy at the VBM. (d) Migration energy profiles for the 33 Ag-related pathways (light blue), comprising $V_{Ag}^{-}$ vacancy hops and $Ag_i^{+}$ interstitialcy paths, with one representative path highlighted, and for the three $V_{Bi}^{2-}$ pathways taken as a framework reference (orange). Each profile shows the twelve movable NEB images and is plotted relative to the lowest image of that same path; the reaction coordinate is the image index normalised to the unit interval.

### 2.2. Switching Evolution in $AgBiS_2$ Memristors

Initially, memristive devices with a simple ITO/Ag-Bi/Ag architecture were fabricated, where the Ag-Bi thickness was ~137 nm (**Figure S1**). The resistive switching characteristics of memristors were evaluated by performing cyclic current-voltage (I-V) sweeps in an optimized voltage range of -1.2 V to +1.2 V over multiple cycles at different current compliance (CC) levels. Memristors with the above device configuration did not exhibit any resistive switching behaviour, as shown in **Figure S2**. This absence of switching is likely due to the presence of pinholes (**Figure S3**), which can create unintended direct conduction paths between the ITO bottom electrode and the silver (Ag) top electrode, thereby preventing controlled filament formation. Therefore, a thin insulating interlayer of Poly(methyl methacrylate) (PMMA) was introduced between the switching layer and the Ag electrode, resulting in an ITO/Ag-Bi/PMMA/Ag structure (**Figure 2a**). These interfacial engineered memristors exhibited clear and stable digital resistive switching characteristics at CC of $I_{CC}$ = 100 mA, with a sharp transition from the high-resistance state (HRS) to the low-resistance state (LRS), referred to as the SET process, occurring at an ultra-low SET voltage ($V_{SET}$) of ~0.10. Furthermore, the devices demonstrated bipolar switching behaviour, with the RESET process occurring under reverse bias at around ~-1.1 V, where the device switched from the LRS back to the HRS. The PMMA layer could play a dual role by eliminating direct electrical contact caused by pinholes and modulating charge injection at the interface, resulting in reliable and stable switching behaviour in these devices.[35] Multiple devices were tested to evaluate reproducibility, and the Ag-Bi devices exhibited consistent SET and RESET voltages.

To achieve better control over the abrupt digital switching behaviour, a charge transport layer (tin oxide: $SnO_2$) was introduced at the bottom electrode (**see Figure 2c**). Interestingly, these devices exhibited gradual (analog) switching instead of digital switching, along with significantly reduced cycle-to-cycle variation compared to devices without the $SnO_2$. The bottom $SnO_2$ interfacial layer likely slows ion migration and promotes interfacial ion accumulation, resulting in gradual switching. Moreover, the $V_{SET}$ shifted to a higher value (~0.54 V) and the $V_{RESET}$ occurred at ~-0.76 V, indicating a more symmetric bipolar switching behaviour. Digital switching is typically required for data storage and logic circuit applications[51], whereas analog switching is desirable for neuromorphic computing.[52] Therefore, this work highlights that through device engineering, the resistive switching mechanism can be effectively tuned and controlled of Ag-Bi based memristors. **Figures 2c** and **2d** present the histograms of the SET and RESET voltages obtained from more than $10^3$ consecutive *I-V* switching cycles for memristors without and with the $SnO_2$ interfacial layer, respectively. For the device without the $SnO_2$ layer, the $V_{SET}$ and $V_{RESET}$ distributions are centred at 0.21 ± 0.019 V and -1.041 ± 0.081 V, respectively. In contrast, the device with the bottom interfacial layer showed more balanced bipolar switching behaviour, with the $V_{SET}$ and $V_{RESET}$ distributions centred at 0.588 ± 0.041 and -0.649 ± 0.09 V, respectively. Long-term cyclic I-V characteristics corresponding to these histograms are presented in the supporting information (**Figure S4**).

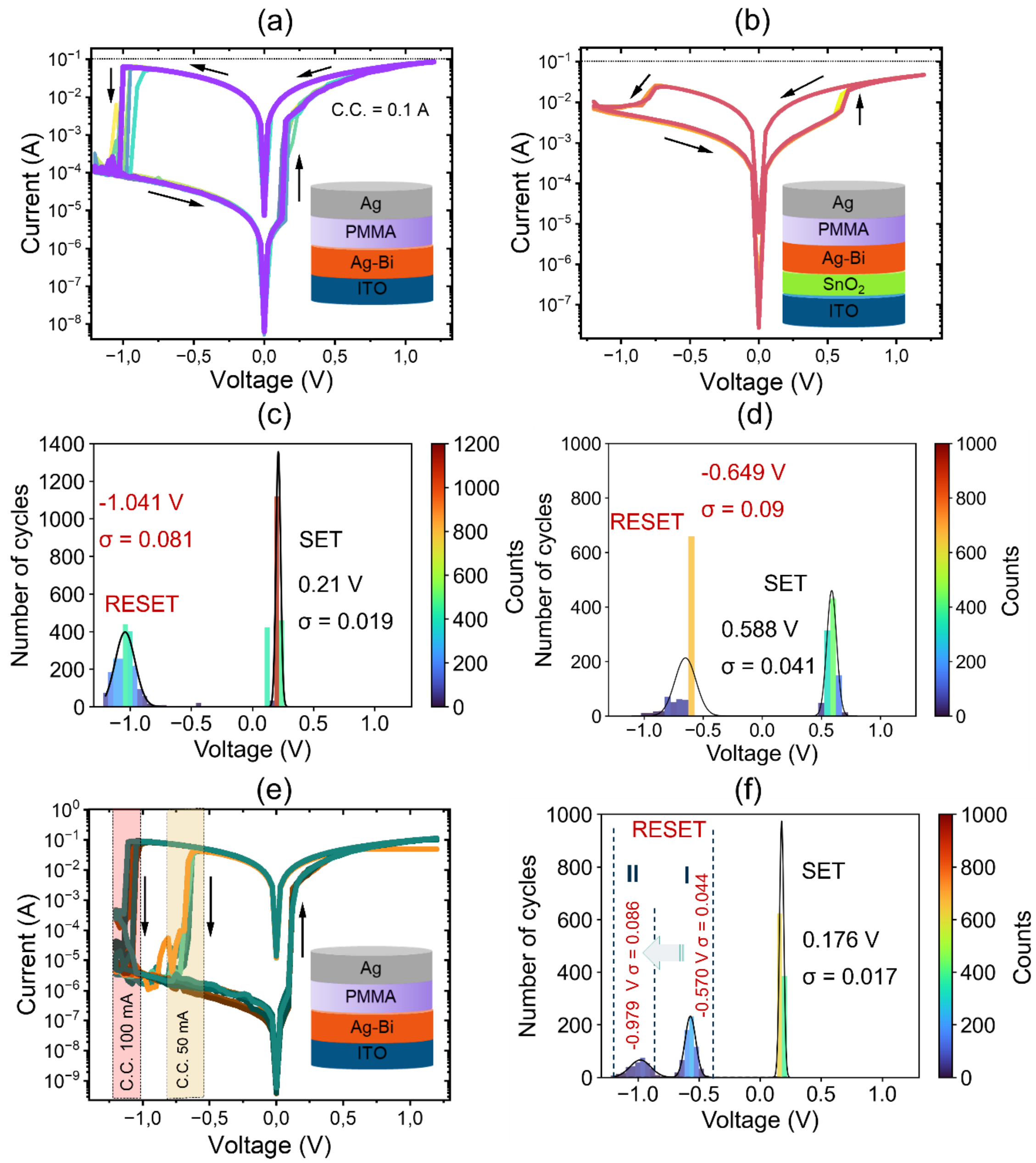

**Figure 2**. *I-V* characteristics of (a) ITO/Ag-Bi/PMMA/Ag and (b) ITO/$SnO_2$/Ag-Bi/PMMA/Ag memristors measured at $I_{CC}$ = 100 mA over the first five cycles within a voltage window of -1.2 to +1.2 V. Histograms of the SET and RESET voltage distributions extracted from more than $10^3$ switching cycles for (c) ITO/Ag-Bi/PMMA/Ag and (d) ITO/$SnO_2$/Ag-Bi/PMMA/Ag devices. (e) Resistive switching characteristics of the ITO/Ag-Bi/PMMA/Ag device measured at $I_{CC}$ = 100 and $I_{CC}$ = 50 mA. (f) Statistical distributions of the SET and RESET voltages obtained from repeated switching cycles at different compliance currents.

Next, the *I-V* characteristics of the memristors were investigated at different $I_{CC}$ levels (**Figure S5a**). While the $V_{SET}$ remained nearly unchanged, the $V_{RESET}$ gradually shifted toward lower values with decreasing $I_{CC}$. **Figure 2e** shows the switching characteristics at $I_{CC}$ = 100 and 50 mA. However, when the CC was reduced below 10 mA, resistive switching was no longer observed (**Figure S5b**), suggesting insufficient conductive filament formation under low-current operation. This behaviour suggests that the initial filament thickness, governed by defect/vacancy distribution and ion migration, along with local Joule heating, plays a crucial role in determining filament formation and stability. Although multiple mechanisms for resistive switching in mixed electronic-ionic materials have been proposed[53], primarily interfacial effects and filament formation, the dominant mechanism in these devices will be discussed in the subsequent sections. To further investigate the influence of CC, long-term cyclic *I-V* measurements were performed at an $I_{CC}$ of 50 mA. As revealed by the distributions of the SET and RESET voltages (**Figure 2f**), no significant change was observed in the $V_{SET}$ values throughout the cycling measurements which is centred at 0.176 ± 0.017 V. In contrast, the $V_{RESET}$ distribution exhibited a pronounced evolution with cycle number. Initially, for more than 500 switching cycles, $V_{RESET}$ remained centred around -0.57 ± 0.044 V. However, upon further cycling, the $V_{RESET}$ distribution evolved from an initial state to a second state centred near -0.979 ± 0.086 V. This behaviour suggests the presence of two distinct switching regimes, with prolonged cycling driving the device toward the more stable regime associated with higher $V_{RESET}$. Interestingly, these results suggest that the switching characteristics induced by lower CC are not permanently programmed. This effect might be attributed from two possible switching kinetics within the $AgBiS_2$ layer. (i) Prolonged cyclic operation at $I_{CC}$ = 50 mA could induce the formation of a secondary, more stable filamentary pathway, which subsequently dominates the switching process (**Figure 3**). (ii) Alternatively, continued cycling may result in the gradual accumulation of mobile interstitial $Ag_i$ ions within the pre-existing filament. Consistent with the first-principles calculations discussed above, which reveal favourable $Ag^+$ migration in $AgBiS_2$, prolonged cycling may progressively reinforce conductive pathways and strengthen the filament. As a result, a higher RESET voltage is required for filament rupture. This interpretation is further supported by the observation that resistive switching persists even after the introduction of the thick PMMA blocking layer, consistent with an intrinsic reservoir of mobile Ag species within the $AgBiS_2$ film.

The CC dependent *I-V* measurements overall suggest that the thickness of the conductive filament formed during the SET process plays a crucial role in the subsequent filament rupture dynamics. At higher CC, a thicker and more robust filament is formed due to enhanced defect migration and ion accumulation; consequently, a higher $V_{RESET}$ is required to rupture the filament during reverse bias. However, reducing the CC leads to the formation of thinner filament, which require less $V_{RESET}$ to break, resulting in a systematic shift of the $V_{RESET}$ toward lower values. Notably, the filament growth does not significantly affect the $V_{SET}$, as SET primarily corresponds to the minimum energy required to establish a conductive path across the switching layer (Ag-Bi). Furthermore, the absence of switching at low CC, indicating the existence of a critical threshold current necessary for stable filament formation. These observations strongly support a filamentary switching mechanism rather than an interfacial one in Ag-Bi based memristors.

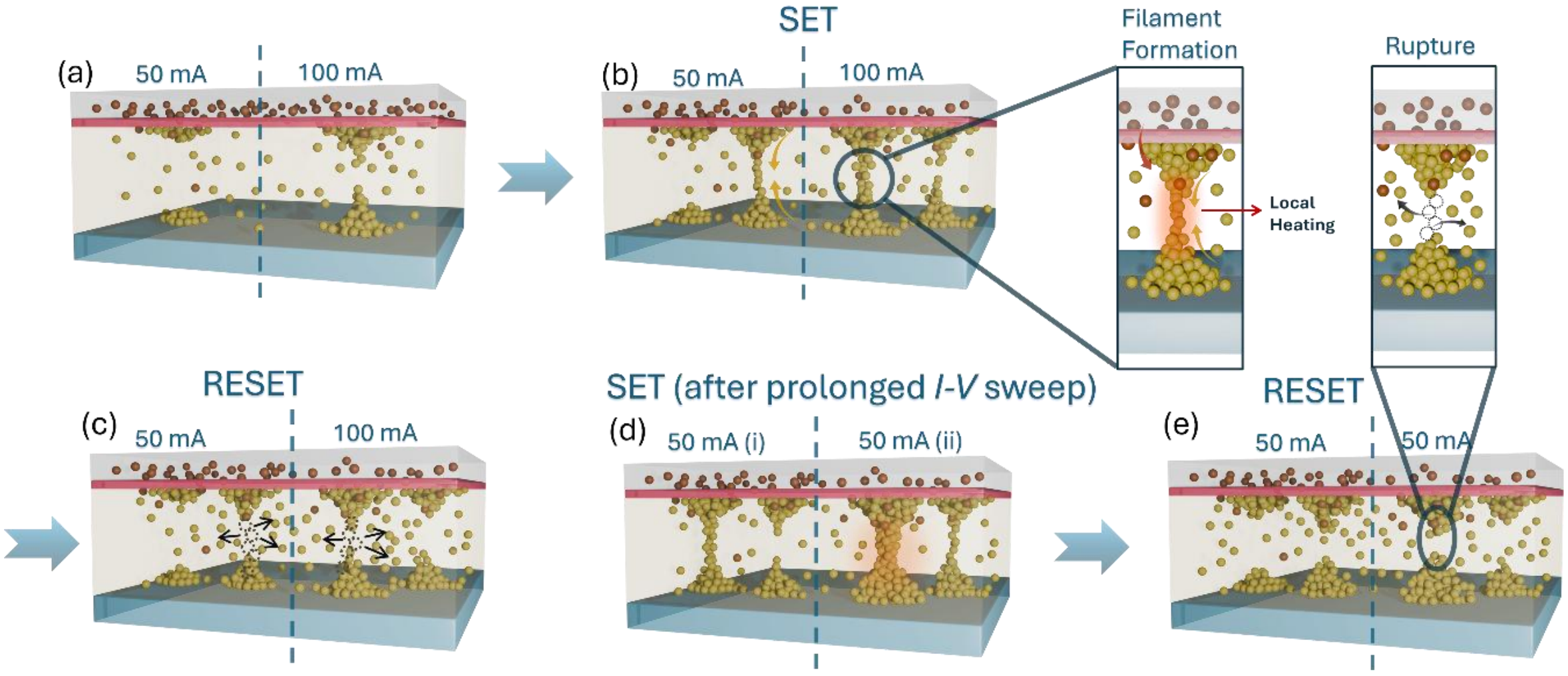

**Figure 3**. Schematic illustration of the resistive switching mechanism in memristors. (a) Random distribution of mobile $Ag^+$ ions/defects in the pristine switching layer. (b) SET process via conductive filament formation involving interstitial and electrode-derived $Ag^+$ ions. (c) RESET process through filament rupture under reverse bias; stronger filaments formed at higher CC require higher $V_{RESET}$. (d) Evolution during prolonged cycling, where (i) filament relocation or (ii) ion accumulation strengthens the filament.

To gain more insight into resistive switching in these memristors, we examined the effect of varying the switching layer thickness on the evolution of resistive switching behaviour. We achieved three distinct Ag-Bi thicknesses d ~ 137 nm, d ~ 235 nm, and d ~ 71 by tuning the spin-coating speed and the number of coating layers, as described in the experimental section (Supporting Information). The *I-V* characteristics discussed so far correspond to devices with a representative thickness of ~137 nm. Notably, memristors with the switching layer of d = 137 nm discussed thus far exhibit forming-free resistive switching from the first cycle (**Figure 2**). However, memristors with thicker switching layer (d = 235 nm) exhibit a distinctly different evolution of resistive switching behaviour (**see Figure 4**). In the initial cycles within a voltage window of ±1.2 V, the devices show unipolar gradual switching with significant hysteresis under positive bias, while no switching and hysteresis are observed under negative polarity (**Figure 4a**). However, in subsequent cycles within the same voltage window, a peak begins to emerge near 0 V (**Figure 4b**), indicating the gradual formation of a conductive pathway. With an increased voltage window, this peak becomes more pronounced, accompanied by a gradual decrease in current around -0.5 V (**Figure 4c**).

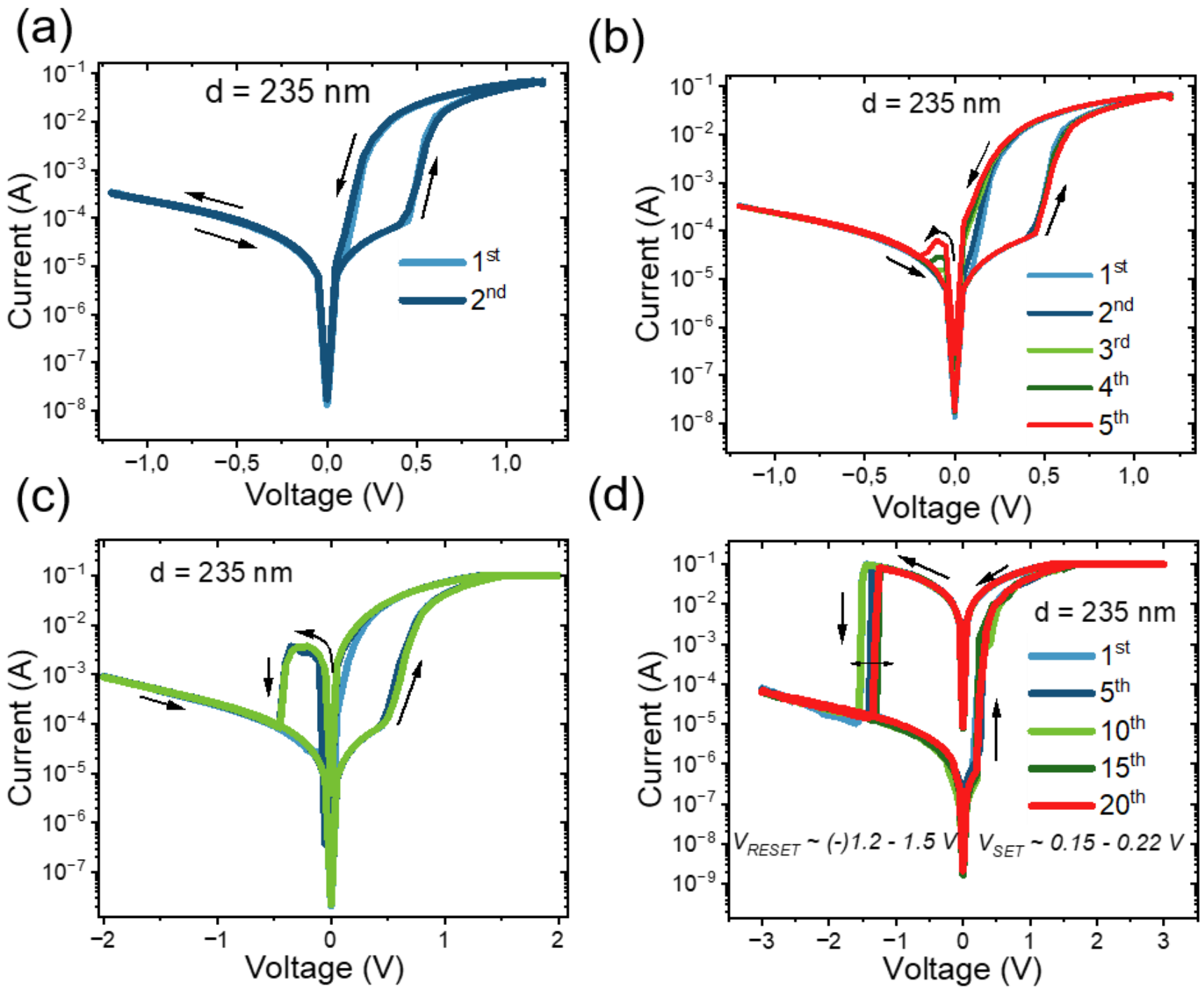

**Figure 4.** Evolution of resistive switching in thick Ag-Bi switching-layer memristors. (a) First two I-V cycles measured within a voltage window of -1.2 to +1.2 V. (b) First five cycles, showing the emergence of a switching peak after the second cycle. (c) Progressive evolution of the switching behaviour with increasing voltage window and cycling, accompanied by a stronger switching peak. (d) Fully developed digital resistive switching exhibiting complete SET and RESET processes at higher voltage windows.

Upon further *I-V* cycling and a larger voltage window of ±3 V, the memristors exhibit a clear abrupt SET process in the range of 0.15 V to 0.22 V and a stable RESET around at -1.2 to -1.5 V (**Figure 4d**), indicating fully developed bipolar resistive switching. Therefore, these thick switching-layer memristors initially displayed gradual unipolar switching, which progressively evolved through metastable states into stable abrupt (digital) switching with repeated cycling and voltage stressing. This programmable evolution can be attributed to a pre-forming process governed by distributed defect clusters. In the initial stage, conduction is likely mediated by the interfacial accumulation of mobile ions and defects, resulting in gradual switching behaviour. With continued cycling, interstitial $Ag^+$ ions, externally supplied $Ag^+$ ions, and intrinsic defects facilitate the formation and dissolution of local defect clusters, giving rise to a metastable switching regime. Finally, upon continued electrical stress, a fully developed filament forms across the switching layer, resulting in stable, abrupt SET and RESET processes.

In thin switching layer memristors exhibited forming-free resistive switching, reliable SET and RESET processes were observed from the first cycle (**Figure S6**). However, the devices exhibited increased current noise and larger cycle-to-cycle variations. The $V_{SET}$ remained in the range of 0.08 - 0.10 V. Notably, the RESET process occurred in two steps: an initial abrupt transition followed by a gradual resistance change. This two-step behaviour might be attributed to the rupture of highly localized conductive filaments, where an initial filament break at the neck region is followed by the gradual dissolution of residual ionic species.

### 2.3.Electronic-Ionic Conduction Mechanisms and Synaptic Properties

To understand the conduction mechanisms in our Ag-Bi memristors, log-log *I-V* characteristics were analysed using the relation $I \propto V^m$.[54] **Figure 5a** depicts the double logarithmic *I-V* plot in first quadrant (positive bias). Here, in the low-voltage region, the slope m = 1.0, indicating Ohmic conduction. At the SET voltage, an abrupt current increase is observed, corresponding to the formation of a conductive filament inside switching layer and transition to the LRS from HRS (Region 1). Notably, in the LRS (Region 2), m = 2.15, suggesting a trap-filled space-charge-limited conduction (SCLC)-type behaviour in the filamentary regime. During the reverse sweep toward 0 V, the slope returns to m = 1, indicating a restoration of Ohmic conduction after partial filament modification. The *I-V* characteristics under negative bias further reveal that the slope remains close to m = 1 until -1.1 V. At this point, a sudden transition occurs, corresponding to the RESET process and rupture of the filament, returning the device to the HRS. After RESET, the conduction again follows an Ohmic behaviour with m = 1 in the HRS. Interestingly, the observed SCLC-like conduction after the SET point in the LRS suggests that the filament is not perfectly metallic and may be influenced by surrounding trap-rich regions in Ag-Bi memristors. Next, the transition to Ohmic region during the reverse sweep indicates a more stabilized and conductive filamentary pathway. From the double logarithmic *I-V* analysis of thinner switching layer (**Figure S6**), it is evident that after the initial abrupt rupture, the conduction follows a high slope (m > 2), indicating that, after partial filament rupture, charge transport is dominated by trap-controlled conduction through the residual filament path. Furthermore, to gain insight into the coupled electronic-ionic dynamics in Ag-Bi, electrochemical impedance spectroscopy (EIS) measurements were performed under varying applied bias in dark conditions. The complex Nyquist spectra, as shown in **Figure 5c** and **Figure S7**, initially exhibit a capacitive response with spectra lying in the first quadrant up to $V_{bias}$= 0.1 V. Notably, beyond this bias, the low-frequency region shifts into the fourth quadrant, forming an inductive loop. Inductive behaviour in mixed electronic-ionic semiconductors is also well-known as apparent negative capacitance, and chemical inductor.[20] The observed inductive behaviour does not correspond to intrinsic negative capacitance or true inductance; rather, it arises from the coupled dynamics of slow ionic conduction and fast electronic transport, which introduce a phase shift in the AC response. The device capacitance was extracted from the EIS spectra, as shown in **Figure 5F**.[20] At low frequency (LF), a high positive capacitance (PC) is observed under low applied bias, which can be attributed to surface polarization at the interfaces.

With increasing applied bias, electronic contributions become dominant, and at a threshold bias, the PC transitions to very high negative capacitance (NC). In our recent work on perovskite solar cells[20], we identified that the transition voltage originates from the competition between ionic conduction and electronic recombination. Based on our previous findings, increased recombination together with slower ionic dynamics shifts the transition voltage to lower values. Therefore, the exceptionally low transition voltage ($V_t$) observed here ($V_t$ > 0.1 V) indicates a high NC response arising from coupled electronic-ionic dynamics in $AgBiS_2$ materials involving interstitial $Ag^+$ ions, Ag vacancies ($V_{Ag}$), sulphur-related defects, and defect-assisted recombination. This anomalous NC or inductive behaviour characterized by intrinsic memory and a delayed response to external stimuli, highlighting their potential for neuromorphic computing applications.[35] This is analogous to that reported for halide perovskites, while offering the additional advantages of environmental compatibility and improved intrinsic stability.

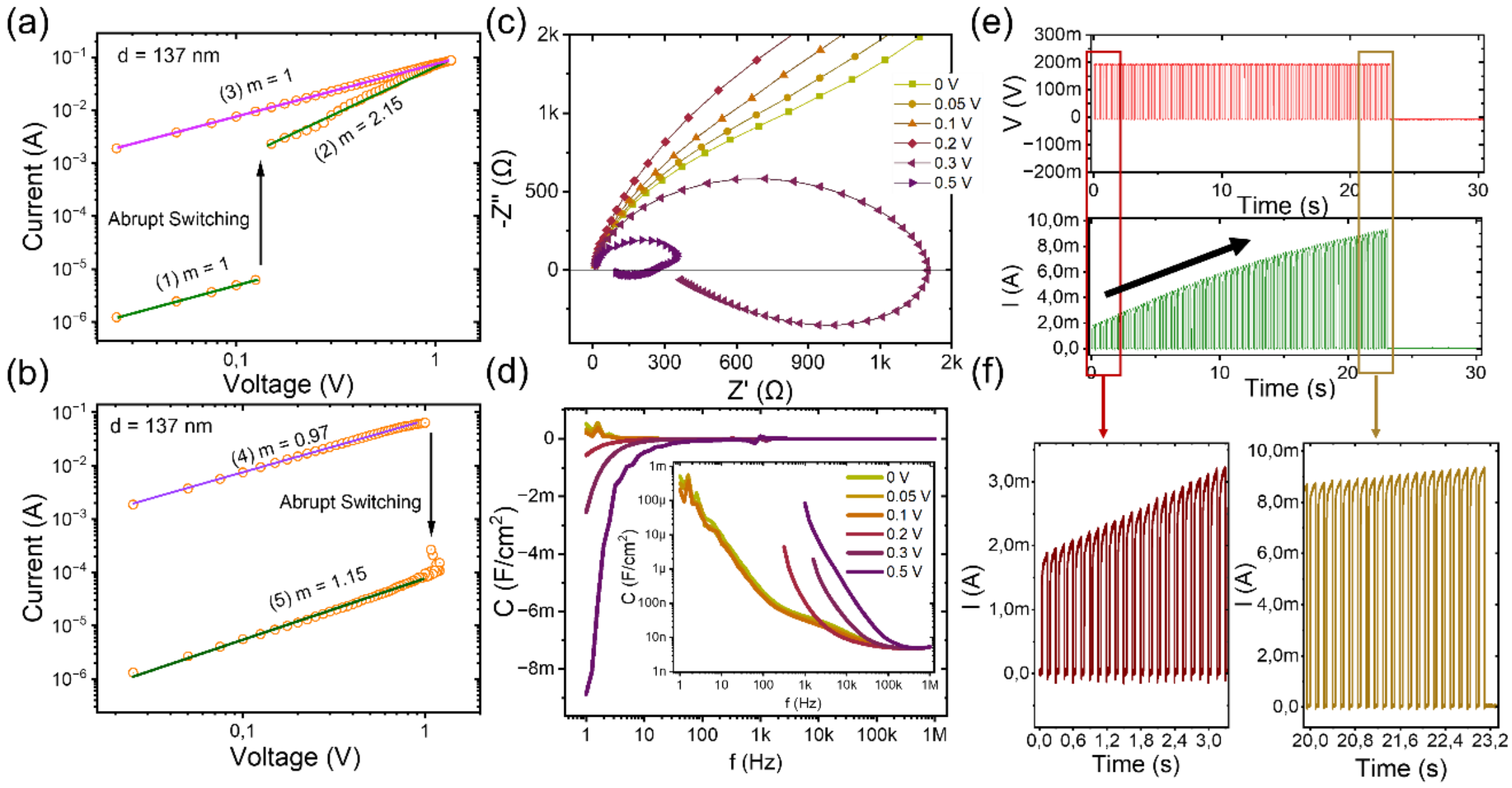

**Figure 5**. Electrical characterization and neuromorphic response of Ag-Bi memristors. (a,b) Log-log *I-V* characteristics under positive and negative bias polarities, respectively. (c) Nyquist spectra at different bias voltages, highlighting the emergence of an inductive loop at low bias. (d) Capacitance-frequency spectra at different bias voltages, showing a pronounced apparent negative capacitance beyond a threshold voltage. (e) Pulse measurements under a voltage pulse train, demonstrating a gradual increase in the output current response. (f) Comparison of the current responses for the initial and final pulses, showing cumulative conductance modulation and synaptic-like behaviour.

To further evaluate their neuromorphic functionality, pulse measurements were performed using a train of 200 mV voltage pulses. As shown in **Figure 5e**, the readout current response to a given pulse depends on the history of preceding pulses. Consequently, the output current increases progressively with the number of applied pulses, exhibiting a cumulative response. Representative current responses corresponding to the initial and final pulses are presented in **Figure 5f**. Such pulse-dependent conductance modulation is analogous to synaptic weight adaptation in biological neural systems, demonstrating the ability of these devices to emulate key synaptic functionalities.

### 2.4. Stability and Structural Evolution of $AgBiS_2$ Memristors

Endurance and retention are also key performance parameters for practical memristor applications. To evaluate memristors stability, we performed continuous endurance measurements over more than 2500 switching cycles for nearly four consecutive days (**Figure 6a**). **Figure 6b** shows selected switching cycles, revealing good cycle-to-cycle reproducibility of the memristors. While the hysteresis magnitude and ON/OFF ratio vary during extended cycling, the devices retain stable and repeatable resistive switching behavior. Furthermore, retention tests confirmed stable preservation of both the HRS and LRS for up to $4\times10^4$ s, demonstrating excellent nonvolatile memory characteristics (**Figure 6c**). Notably, the memristors remained operational after the long *I-V* measurements. To probe possible structural modifications associated with switching, the devices were subsequently cleaved and characterized using cross-sectional SEM. As shown in **Figure 6d** (top), the pristine switching layer is compact and uniform. In contrast, the aged device (**Figure 6d, bottom**) exhibits noticeable swelling of the switching layer, and the regions highlighted by circles suggest local material fragmentation. These structural modifications are likely associated with repeated filament formation and rupture during resistive switching. The observed morphological evolution indicates that, although stable SET and RESET operations are maintained, continuous structural rearrangements within the switching layer contribute to cycle-to-cycle variations in the device characteristics. This variability may also be intrinsically linked to the cation-disordered nature of $AgBiS_2$. As revealed by our first-principles calculations, the local atomic environment gives rise to a broad distribution of defect energetics and migration barriers, spanning several tenths of an eV, which may in turn underlie the stochastic nature of filament formation and rupture during resistive switching. Together, these structural and defect-mediated effects provide a plausible microscopic origin for the observed cycle-to-cycle variability. Therefore, future efforts should focus on minimizing structural degradation and achieving more controlled filament formation and rupture to improve the long-term stability and reliability of these memristors.

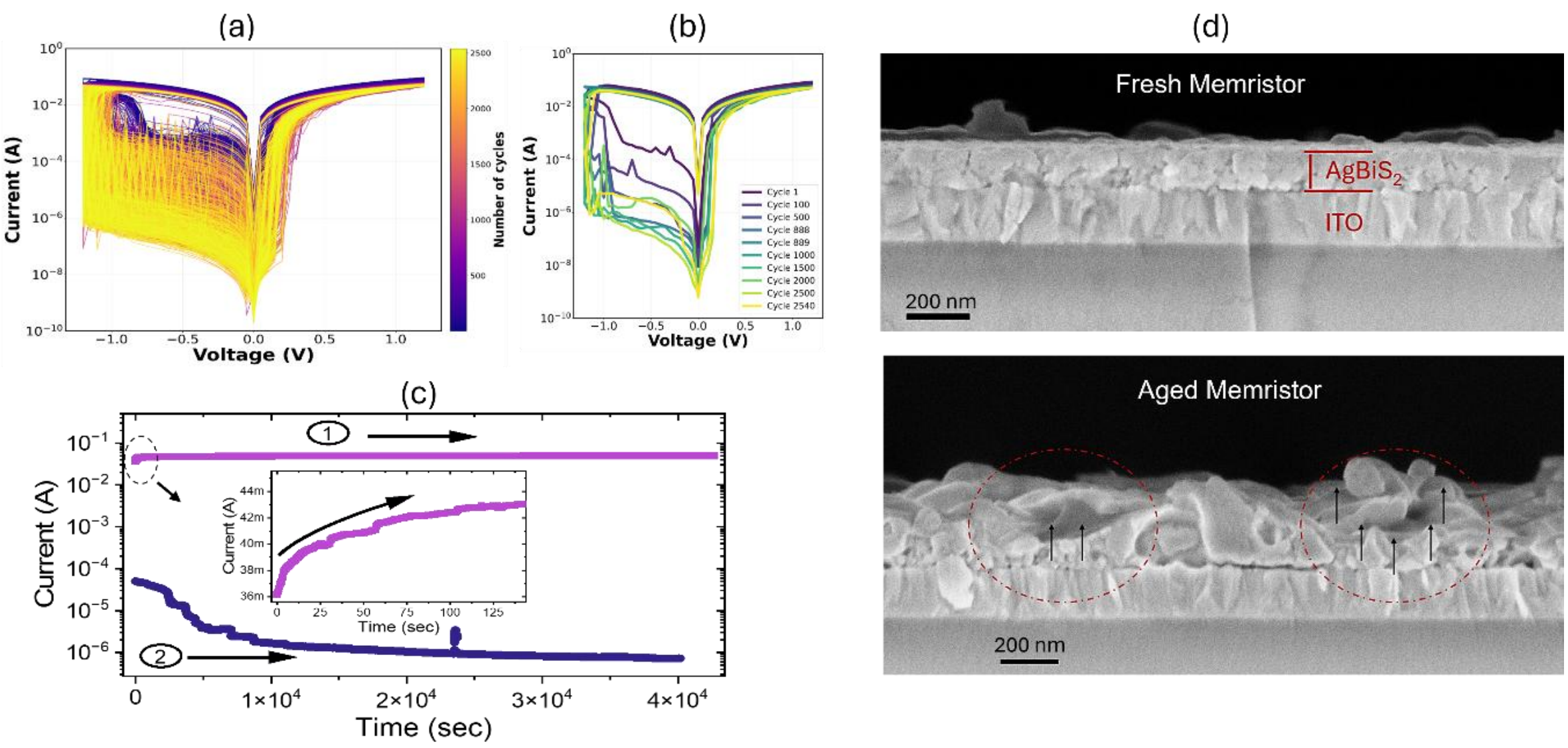

**Figure 6**. (a) Endurance performance of the Ag-Bi memristors over more than 2500 consecutive switching cycles. (b) I-V chrematistics extracted from selected cycles. (c) Retention characteristics of the LHS and HRS measured over extended periods. (d) Cross-

sectional SEM image of a pristine memristor. (e) Cross-sectional SEM image of an aged device after prolonged cyclic switching, revealing structural modifications within the switching layer.

## 3. Conclusion

In summary, perovskite-inspired chalcogenide memristors with engineered interfaces and controlled switching-layer thickness exhibit tunable switching behaviour ranging from analog and metastable states to fully developed digital switching. The devices demonstrate an ultra-low SET voltage (~0.08 V), high ON/OFF ratios, and forming-free operation. Combined electrical measurements and first-principles calculations reveal that defect-assisted $Ag^+$ migration and filament evolution govern the resistive switching process. Electrochemical impedance spectroscopy highlights strong coupled electronic-ionic dynamics manifested as pronounced apparent negative capacitance, which is further reflected in the observed synaptic-like responses under pulse stimulation. Long-term stability and structural characterization reveal that repeated switching induces local morphological changes within the switching layer, highlighting the importance of controlling filament dynamics to improve device reliability. Overall, these findings establish $AgBiS_2$ as a promising lead-free, solution-processable memristive platform and provide design strategies for realizing energy-efficient memory and neuromorphic computing technologies.

## 4. Experimental Section

Full details of experimental procedures can be found in the Supplemental Information

Data and Code Availability

All supporting results are provided in the Supplemental Information. For further data requests, please contact the Lead Contact (Ramesh Kumar).

**Supplemental Information**

Supporting Information is available.

**Funding Sources**

This work primarily supported by the Young Investigator Group Preparation Program, funded jointly through the University of Excellence Strategy Fund at the Karlsruhe Institute of Technology (KIT) and the Ministry of Science, Research, and the Arts of Baden-Württemberg, Germany.

**Author Contributions**

R. K. contributed to the conceptualization, design of the study, data analysis, writing original draft, funding acquisition and supervision. E.J.S, L.Z., and J.M., fabricated devices, performed all device measurements, and helped in writing. S.D. optimized the $AgBiS_2$ precursor formulation and thin-film deposition process and contributed to device fabrication. Z.L. and V.C. designed and performed simulation work, data interpretation, and writing. B.R. performed XRD, and SEM measurements. A.S. performed data interpretation. R. K., G.H.S, M.B. and H.Z. were responsible for supervision. All authors contributed to the manuscript draft and editing.

**ACKNOWLEDGMENTS**

R.K. acknowledges support from the Young Investigator Group Preparation Program, funded jointly through the University of Excellence Strategy Fund at the Karlsruhe Institute of Technology (KIT) and the Ministry of Science, Research, and the Arts of Baden-Württemberg, Germany. G.H.S. thanks the German Research Foundation DFG for the financial support via the Heisenbergprofessur, HE 7056/7-1. S.D and G.H.S acknowledge funding by the Carl Zeiss Foundation (project KeraSolar). We acknowledge Myfab Uppsala for providing facilities and experimental support. Myfab is funded by the Swedish Research Council (2020-00207) as a national research infrastructure.

**Declaration of interests**

The authors declare no conflict of interest.

**References**

[1] C. Mead, *Proc. IEEE* **1990**, *78*, 1629.

[2] A. Gholami, Z. Yao, S. Kim, C. Hooper, M. W. Mahoney, K. Keutzer, *IEEE Micro* **2024**, *44*, 33.

[3] K. Roy, A. Kosta, T. Sharma, S. Negi, D. Sharma, U. Saxena, S. Roy, A. Raghunathan, Z. Wan, S. Spetalnick, C.-K. Liu, A. Raychowdhury, *Front. Sci.* **2025**, *3*, 1611658.

[4] S. Yu, *Proc. IEEE* **2018**, *106*, 260.

[5] D. Kudithipudi, C. Schuman, C. M. Vineyard, T. Pandit, C. Merkel, R. Kubendran, J. B. Aimone, G. Orchard, C. Mayr, R. Benosman, J. Hays, C. Young, C. Bartolozzi, A. Majumdar, S. G. Cardwell, M. Payvand, S. Buckley, S. Kulkarni, H. A. Gonzalez, G. Cauwenberghs, C. S. Thakur, A. Subramoney, S. Furber, *Neuromorphic computing at scale*, Vol. 637, Nature Publishing Group, **2025**, pp. 801–812.

[6] H. Greatorex, O. Richter, M. Mastella, M. Cotteret, P. Klein, M. Fabre, A. Rubino, W. Soares Girão, J. Chen, M. Ziegler, L. Bégon-Lours, G. Indiveri, E. Chicca, *Nat. Commun.* **2025**, *16*, 1.

[7] D. Das, D. S. Assi, S. Kazim, V. A. L. Roy, S. Ahmad, *Decoding halide perovskites for neuromorphic and memristive devices*, Vol. 12, The Royal Society of Chemistry, **2025**, pp. 8430–8459.

[8] Q. Chen, L. Lu, J. Meng, M. Xu, T. Wang, *Research* **2025**.

[9] D. Hasina, A. Mandal, S. K. Srivastava, A. Mitra, T. Som, *Small* **2025**, *21*, 2408369.

[10] S. Rani, P. Kaith, A. Ghadge, S. A. Siddiqui, S. Das, M. Afshan, D. Rani, N. Chaudhary, E. M. Harini, M. Bag, A. Bera, K. Ghosh, *Adv. Funct. Mater.* **2026**, e77208.

[11] X. Yin, Y. Wang, T. Chang, P. Zhang, J. Li, P. Xue, Y. Long, J. L. Shohet, P. M. Voyles, Z. Ma, X. Wang, *Adv. Mater.* **2020**, *32*, 2000801.

[12] J. Zheng, R. Wang, W. Fang, C. Li, J. Zhang, Z. Wan, Y. Chen, J. Liu, X. Zou, L. Xie, Q. Wang, X. Li, S. Song, X. Zhou, Z. Song, *Nat. Commun.* **2025**, *16*, 5788.

[13] J. Choi, J. S. Han, K. Hong, S. Y. Kim, H. W. Jang, *Adv. Mater.* **2018**, *30*, 1704002.

[14] S. J. Kim, H. J. Lee, G. B. Nam, J. Y. Kim, H. W. Jang, *Halide Perovskites for Neuromorphic Sensing and Computing*, Vol. 17, American Chemical Society, **2025**, pp. 59951–59978.

[15] N. Yantara, N. Mathews, *Toolsets for assessing ionic migration in halide perovskites*, Vol. 8, Cell Press, **2024**, pp. 1239–1273.

[16] M. A. Green, A. Ho-Baillie, H. J. Snaith, *The emergence of perovskite solar cells*, Vol. 8, Nature Publishing Group, **2014**, pp. 506–514.

[17] Y. H. Kim, H. Cho, T. W. Lee, *Proc. Natl. Acad. Sci. U. S. A.* **2016**, *113*, 11694.

[18] N. Li, Y. Jia, Y. Guo, N. Zhao, *Adv. Mater.* **2022**, *34*, 2108102.

[19] J. Thiesbrummel, J. V. Milić, C. Deibel, E. C. Garnett, S. Tao, T. Kirchartz, A. Guerrero, P. Cameron, W. Tress, M. Saiful Islam, B. Ehrler, *Ion migration in perovskite solar cells*, Vol. 10, Nature Publishing Group, **2026**, pp. 179–195.

[20] R. Kumar, B. Rakheja, N. Lamminen, F. Fasulo, M. A. T. Cachafeiro, C. Hanmandlu, G. K. Grandhi, M. Bag, A. B. Muñoz-García, G. Boschloo, W. Tress, M. Pavone, P. Vivo, E. M. J. Johansson, *Adv. Energy Mater.* **2025**, *15*, e03331.

[21] W. Tress, N. Marinova, T. Moehl, S. M. Zakeeruddin, N. Mohammad K., M. Grätzel, M. K. Nazeeruddin, M. Grätzel, *Energy Environ. Sci.* **2015**, *8*, 995.

[22] F. Ebadi, N. Taghavinia, R. Mohammadpour, A. Hagfeldt, W. Tress, *Nat. Commun.* **2019**, *10*.

[23] E. Ghahremanirad, A. Bou, S. Olyaee, J. Bisquert, *J. Phys. Chem. Lett.* **2017**, *8*, 1402.

[24] J. Bisquert, *PRX Energy* **2024**, *3*, 11001.

[25] E. Hernández-Balaguera, J. Bisquert, *Adv. Funct. Mater.* **2024**, *34*, 2308678.

[26] J. Bisquert, A. Guerrero, *J. Am. Chem. Soc.* **2022**, *144*, 5996.

[27] S. E. Ng, N. Yantara, N. Mathews, *Halide Perovskite Retinomorphic In-Sensor Computing*, Vol. 10, American Chemical Society, **2025**, pp. 5771–5780.

[28] K. Dedecker, G. Grancini, K. Dedecker, G. Grancini, *Adv. Energy Mater.* **2020**, *10*, 2001471.

[29] V. K. Ravi, B. Mondal, V. V. Nawale, A. Nag, *Don't let the lead out: New material chemistry approaches for sustainable lead halide perovskite solar cells*, Vol. 5, American Chemical Society, **2020**, pp. 29631–29641.

[30] R. Vidal, N. Lamminen, V. Holappa, J. A. Alberola-Borràs, I. P. Franco, G. K. Grandhi, P. Vivo, *Adv. Energy Mater.* **2025**, *15*, 2403981.

[31] G. K. Grandhi, D. Hardy, M. Krishnaiah, B. Vargas, B. Al-Anesi, M. P. Suryawanshi, D. Solis-Ibarra, F. Gao, R. L. Z. Hoye, P. Vivo, *Adv. Funct. Mater.* **2024**, *34*, 2307441.

[32] Y. Peng, T. N. Huq, J. Mei, L. Portilla, R. A. Jagt, L. G. Occhipinti, J. L. MacManus-Driscoll, R. L. Z. Hoye, V. Pecunia, *Adv. Energy Mater.* **2021**, *11*, 2002761.

[33] Noora Lamminen, Joshua Karlsson, Ramesh Kumar, N. S. Manikanta Viswanath, Snigdha Lal, Francesca Fasulo, Marcello Righetto, Mokurala Krishnaiah, Kimmo Lahtonen, Amit Tewari, Atanas Katerski, Jussi Lahtinen, I. O. Acik, E. M. J. Johansson, A. Belén Muñoz-García, Michele Pavone, L. M. Herz, G. Krishnamurthy Grandhi, Paola Vivo, *EES Sol.* **2025**, *1*, 139.

[34] D. B. Mitzi, *Inorg. Chem.* **2000**, *39*, 6107.

[35] R. Kumar, N. Lamminen, A. Ghadge, M. Bag, G. K. Grandhi, P. Vivo, *Small Struct.* **2026**, *7*, e202500842.

[36] J. He, X. Hu, Z. Liu, W. Chen, G. Longo, *Prospect for Bismuth/Antimony Chalcohalides-Based Solar Cells*, Vol. 33, John Wiley and Sons Inc, **2023**, p. 2306075.

[37] Y. Zhu, J. S. Liang, V. Mathayan, T. Nyberg, D. Primetzhofer, X. Shi, Z. Zhang, *ACS Appl. Mater. Interfaces* **2022**, *14*, 21173.

[38] H. Sun, Q. Lu, F. Yang, X. Zhang, X. Dong, J. Chen, X. Zhang, J. Chen, Y. Zhao, Y. Li, *Adv. Funct. Mater.* **2026**, *36*, e75808.

[39] W. Yang, T. Sun, H. Yu, H. Shi, Y. Hu, J. Huang, Z. Liu, Y. Xu, L. Wang, B. Hu, Y. Shen, M. K. Nazeeruddin, M. Wang, *Nat. Commun. 2026 171* **2026**, *17*, 5687.

[40] M. Bernechea, N. C. Miller, G. Xercavins, D. So, A. Stavrinadis, G. Konstantatos, *Nat. Photonics* **2016**, *10*, 521.

[41] S. Akhil, R. G. Balakrishna, *AgBiS2 as a photoabsorber for eco-friendly solar cells: a review*, Vol. 10, The Royal Society of Chemistry, **2022**, pp. 8615–8625.

[42] M. Mohammadi, T. Sachsenweger, E. L. Comi, F. Ji, S. Lohde, M. A. Torre Cachafeiro, A. K. Sachan, K. P. Pernstich, E. Knapp, W. Tress, *Joule* **2026**, *0*.

[43] W. Tress, *Metal Halide Perovskites as Mixed Electronic-Ionic Conductors: Challenges and Opportunities - From Hysteresis to Memristivity*, Vol. 8, American Chemical Society, **2017**, pp. 3106–3114.

[44] H. Sun, Q. Lu, F. Yang, X. Zhang, X. Dong, J. Chen, X. Zhang, J. Chen, Y. Zhao, Y. Li,

*Adv. Funct. Mater.* **2026**, *36*, e75808.

[45] H. Sharma, N. Saini, A. Kumar, R. Srivastava, *J. Mater. Chem. C* **2023**, *11*, 11392.

[46] A. Zunger, S. H. Wei, L. G. Ferreira, J. E. Bernard, *Phys. Rev. Lett.* **1990**, *65*, 353.

[47] Y. Ji, Q. Zhong, X. Yang, L. Li, Q. Li, H. Xu, P. Chen, S. Li, H. Yan, Y. Xiao, F. Xu, H. Qiu, Q. Gong, L. Zhao, R. Zhu, *Nano Lett.* **2024**, *24*, 10418.

[48] S. P. Ong, W. D. Richards, A. Jain, G. Hautier, M. Kocher, S. Cholia, D. Gunter, V. L. Chevrier, K. A. Persson, G. Ceder, *Comput. Mater. Sci.* **2013**, *68*, 314.

[49] J.-X. Shen, J. Varley, *J. Open Source Softw.* **2024**, *9*, 5941.

[50] G. Henkelman, H. Jónsson, *J. Chem. Phys.* **2000**, *113*, 9978.

[51] J. Borghetti, G. S. Snider, P. J. Kuekes, J. J. Yang, D. R. Stewart, R. S. Williams, *Nature* **2010**, *464*, 873.

[52] R. Deb, F. Yasmin, Y. Mishra, Z. Azmi, D. Sahoo, S. R. Mohapatra, *RSC Appl. Interfaces* **2026**, *3*, 168.

[53] W. Sun, B. Gao, M. Chi, Q. Xia, J. J. Yang, H. Qian, H. Wu, *Understanding memristive switching via in situ characterization and device modeling*, Vol. 10, Nature Publishing Group, **2019**, pp. 3453-.

[54] S. Y. Kim, J. Bisquert, *Adv. Funct. Mater.* **2026**, *36*, e75888.